\documentclass[twocolumn,english,nofootinbib]{revtex4}
\usepackage[T1]{fontenc}
\usepackage[latin9]{inputenc}
\usepackage[normalem]{ulem}
\usepackage{color}
\usepackage{babel}
\usepackage{amsmath}
\usepackage{color}
\usepackage{graphicx}
\usepackage{esint}
\usepackage[unicode=true,
 bookmarks=true,bookmarksnumbered=false,bookmarksopen=false,
 breaklinks=false,pdfborder={0 0 0},backref=false,colorlinks=false]
 {hyperref}

\newcommand{\be}{\begin{equation}}
\newcommand{\ee}{\end{equation}}
\newcommand{\nn}{\nonumber}

\hypersetup{pdftitle={New Geometric Representations of the CMB 2pcf},
 pdfauthor={A. Froes et. al.},
 pdfkeywords={CMB, Statistical Isotropy}}

\makeatletter
\@ifundefined{textcolor}{}
{%
 \definecolor{BLACK}{gray}{0}
 \definecolor{WHITE}{gray}{1}
 \definecolor{RED}{rgb}{1,0,0}
 \definecolor{GREEN}{rgb}{0,1,0}
 \definecolor{BLUE}{rgb}{0,0,1}
 \definecolor{CYAN}{cmyk}{1,0,0,0}
 \definecolor{MAGENTA}{cmyk}{0,1,0,0}
 \definecolor{YELLOW}{cmyk}{0,0,1,0}
}

\makeatother

\begin{document}

\title{New Geometric Representations of the CMB 2pcf}

\author{Andre L. D. Froes, Thiago S. Pereira}
\email[Corresponding author: ]{tspereira@uel.br}
\affiliation{Departamento de Física, Universidade Estadual de Londrina, Rodovia
Celso Garcia Cid, km 380, 86051-990, Londrina -- PR, Brazil}

\author{Armando Bernui}
\affiliation{Observatório Nacional, Rua General José Cristino 77, São Cristóvão,
20921-400, Rio de Janeiro -- RJ, Brazil}

\author{Glenn D. Starkman}
\affiliation{CERCA, Physics Department, Case Western Reserve University, Cleveland
-- OH, 44106-7079, USA}

\begin{abstract}
When searching for deviations of statistical isotropy in CMB, a popular
strategy is to write the two-point correlation function (2pcf) as
the most general function of four spherical angles (i.e., two unit
vectors) in the celestial sphere. Then, using a basis of bipolar spherical
harmonics, statistical anisotropy will show up if and only if any
coefficient of the expansion with non-trivial bipolar momentum is detected
-- although this detection will not in general elucidate the origin
of the anisotropy. In this work we show that two new sets of four angles
and basis functions exist which completely specifies the 2pcf, while,
at the same time, offering a possible geometrical interpretation of
the mechanisms generating the signal. Since the coefficients of these
expansions are zero if and only if isotropy holds, they act as a simple 
and geometrically motivated null test of statistical isotropy, with the advantage 
of allowing cosmic variance to be controlled in a systematic way. We report the 
results of the application of these null tests to the latest temperature data released
by the Planck collaboration.
\end{abstract}
\maketitle

\section{Introduction}

Given the unprecedented limits on cosmological parameters achieved
with {\it Planck} data releases~\cite{Ade2014d,Ade2015a}, an important
follow-up question is whether the same data contain traces of physics
beyond the standard $\Lambda$CDM model. While deep field 
surveys~\cite{Laureijs2011,Abate2012,Benitez2014} aim to unveil the specific
nature of dark matter and dark energy, and thus of the energy content
of our universe, CMB observations are special in the sense that they
provide a unique window to both the physics of the early inflationary
universe and of its global shape, i.e., its geometry and 
topology~\cite{Cornish2004,Kunz2006,Ade2015,Fabre:2013wia}. 

From the statistical point of view, finding evidences of new features of inflationary
physics or of the shape of the universe usually translates into cosmological
detections of non-Gaussianity and statistical anisotropy, respectively (but not 
necessarily~\cite{Schmidt:2012ky}).
As it turns out, however, observational bounds from CMB on 
non-Gaussianity~\cite{Komatsu2003,Ade2014c,Ade2015b} did not allow us to discriminate
inflationary models, and the hope now lies on the possibility that
future measurements of polarization $B$-modes of CMB~\cite{Bouchet2011}
will better elucidate the physics of the early universe. The physics
describing the global topology and geometry of the universe, on the other hand, is
not only much less constrained by CMB \cite{Ade2015}, but is also
equally fundamental to the beyond-$\Lambda$CDM program. In this
regard, the existence of statistical anomalies at the largest CMB
angles (see Ref. \cite{Copi2010} for a comprehensive review) can
be optimistically seen as an indication of spatial anisotropy at
the horizon scales \cite{Pitrou2008}, although the conservative minded
would also remind us of possible unaccounted systematic effects \cite{Hanson2010a},
or even the less exciting case of statistical flukes \cite{Bennett2011}. 

Regardless of the final words on CMB anomalies, however, one is rightfully justified to question the 
validity of the statistically isotropic scenario, given its deep connections with the symmetry 
hypothesis about our universe. Thus motivated, this paper addresses the question of how to constrain 
deviations from statistical isotropy in a geometrically meaningful and model-independent way.

The general recipe for describing the statistics of a {\it Gaussian} and statistically isotropic CMB map 
is straightforward. Assuming that the geometry of the universe is everywhere rotationally invariant, 
all we need to do is to compute the correlation between the temperature fluctuations $\Delta T$
at directions $\hat{n}$ and $\hat{n}'$, given by
\begin{equation}
C(\hat{n},\hat{n}')=\left.\left\langle \Delta T(\hat{n})
\Delta T(\hat{n}')\right\rangle \right|_{\cos\theta=\hat{n}\cdot\hat{n}^{\prime}}
\equiv C(\theta)\,.\label{iso-2pcf}
\end{equation}
However, if there are departures from statistical isotropy, either of systematic~\cite{Hanson2010a}, 
astrophysical~\cite{Yoho01072013,Ade2014b,Amendola2011} or cosmological 
\cite{Pitrou2008,Kahniashvili2008a,BH08,Durrer1998} origin, the two-point correlation function 
(2pcf) will depend on other angles relating $\hat{n}$ and $\hat{n}'$; in this case, $C(\theta)$ no 
longer exhausts the statistics of the universe. Thus, if we want to go beyond the 
$\Lambda$CDM framework, the central question is how to parameterize deviations from 
Eq.~(\ref{iso-2pcf}) in a meaningful, and hopefully practical, way.

The angular correlation function has come under considerable scrutiny.  It was first noticed by the 
COBE-DMR team \cite{Hinshaw:1996ut} that $C(\theta)$ was unexpectedly close to zero for 
$\theta\gtrsim60^\circ$.   This lack of correlations was confirmed by the Wilkinson Microwave Anisotropy 
Probe (WMAP) team in their analysis of their first year of data \cite{Spergel2003}.  Though WMAP 
claimed to have greatly reduced significance in future releases,  Copi {\it et al.}  showed 
\cite{Copi:2006tu} that in fact this absence of large-angle correlations persisted on the sky 
outside the galaxy in the third year release, and in all subsequent releases, including the 
first-year Planck release \cite{Copi:2008hw, Copi:2010na,Copi:2013cya}. Those findings have since 
been confirmed by others \cite{Hajian:2007pi,Bunn:2008zd}, although no satisfactory explanation 
exists. It has been suggested \cite{Copi:2013zja,Yoho:2013tta,Yoho:2015bla} that one might be able 
to test whether the anomalous vanishing  of the temperature-temperature correlation function 
is due to new physics or just a statistical anomaly by examining other  two-point correlation 
functions (eg. the $E$-mode-$E$-mode correlation function) on similarly large angles.

A common strategy for parameterizing deviations of isotropy in the CMB is to use a complete set of 
basis functions to perform a multipolar expansion of the 2pcf. Since the latter is defined by the 
product of two functions on the CMB sphere, we could simply use the product of two independent 
spherical harmonics as such basis. Instead, it has become a standard practice to use a basis of 
total angular momentum eigenfunctions, also known as bipolar spherical 
harmonics~\cite{Varshalovich1988a}, to do the expansion. Besides sharing most of the mathematical 
properties of the standard spherical harmonics, the advantage of the bipolar harmonics is that they 
encode deviations of isotropy in the total angular momentum of the coefficients of the expansion. 
Thus, any measurement of a multipolar coefficient with a non-trivial total angular momentum 
eigenvalue is an indication of statistical anisotropy. This program was introduced by Hajian and 
Souradeep~\cite{Hajian2003a,Hajian2004,Hajian2005}, and has been fruitfully applied to 
CMB since then~\cite{Joshi2010,Ade2014b,Book2012,Kamionkowski2011,Hajian2006}. 

Since the bipolar spherical harmonics form a basis for square-integrable functions on the Hilbert 
space where the 2pcf is defined, they offer a very general framework for studying the statistics of 
the CMB. However, it has some limitations, too. First, the multipolar coefficients of the bipolar 
expansion, while serving as a null indicator of anisotropy, do not provide a physical interpretation
of the underlying signal straightforwardly, and for that one has to resort to other tools 
\cite{Kumar2014}. Second, in a more symmetric situation -- whether real or expected -- it is
not clear how the degrees of freedom of the bipolar spherical harmonics
could be combined to reduce cosmic variance.

This paper is based on the idea that, given two unit vectors rooted at the origin of the CMB 
sphere, two new and \emph{unique} geometrical objects can be formed: a great circle and (the 
boundary circle of) a spherical cap -- or, if we include the interior of the sphere, a disc and a cone. 
Using the set of angles defined by these objects, one can introduce a complete set of basis
functions that characterize the 2pcf, and whose multipolar coefficients offer a direct and 
geometrically motivated null test of statistical isotropy. Moreover, since the angles used to 
represent the 2pcf have a clear geometrical interpretation, it may at times be physically 
well-motivated to integrate, average or marginalize over one of them.

This work is organized as follows. After reviewing the basics of the bipolar spherical harmonic 
formalism in Sec. \ref{sub:biposh}, we introduce the new geometric representations of the 2pcf in 
Secs. \ref{sub:cone} and \ref{sub:disc}, where we also show how they recover the usual
2pcf in the isotropic limit. In Sec. \ref{sec:null-tests} we show how these new functions can be 
used to construct null geometrical tests of statistical isotropy. Finally, we present the results 
of null tests applied to the latest temperature maps released by the {\it Planck} team in Sec. 
\ref{sub:data}. We conclude in Sec. \ref{sec:conclusions} with some 
perspectives of future developments.

\section{Geometric Representations}

We start by recalling the basic definitions and notations used in
CMB statistics. The temperature fluctuation field is a real function
on the sphere, and can thus be expanded as
\[
\Delta T(\hat{n})=\sum_{l,m}a_{lm}Y_{lm}(\hat{n})\,,
\]
where $\hat{n}=(\theta,\phi)$. 
In the canonical $\Lambda$CDM cosmological model, the $a_{lm}$ are
realizations of a Gaussian-random and statistically independent variables.
In this case the expectation value of the two-point correlation function is
\begin{equation}
C\left(\hat{n}_{1},\hat{n}_{2}\right)=\!\!\sum_{\substack{l_{1},m_{1}\\
l_{2},m_{2}}}\!\!
\left\langle a_{l_{1}m_{1}}a^*_{l_{2}m_{2}}\right\rangle Y_{l_{1}m_{1}}(\hat{n}_{1})Y^*_{l_{2}m_{2}}(\hat{n}_{2})\,.
\label{general-2pcf}
\end{equation}
The set of coefficients
$\left\langle a_{l_{1}m_{1}}a^*_{l_{2}m_{2}}\right\rangle $ form the
\emph{covariance matrix}, and in the canonical statistically isotropic case
\begin{equation}
\left\langle a_{l_{1}m_{1}}a_{l_{2}m_{2}}^{*}\right\rangle =C_{l_{1}}\delta_{l_{1}l_{2}}\delta_{m_{1}m_{2}}\,.\label{SI}
\end{equation}
Any non-zero off-diagonal term in the covariance matrix is a measure of statistical anisotropy.%
\footnote{Conversely, a diagonal matrix \emph{does not} imply isotropy, since $C_{l_{1}}$ could 
depend on $m_{1}$. Statistical isotropy is thus a strong condition requiring \emph{both} 
independence among multipoles and the invariance of $C_{l_{1}}$ by rotations.}

The 2pcf is symmetric by definition
\begin{equation}
C(\hat{n}_{1},\hat{n}_{2})=C(\hat{n}_{2},\hat{n}_{1})\,.\label{exchange-symmetry}
\end{equation}
As we shall see, this symmetry imposes restrictions on the eigenvalues of the eigenfunctions (and 
consequently on the multipolar coefficients) that we introduce below.

\subsection{Bipolar representation\label{sub:biposh}}

A convenient basis for expanding the 2pcf is given by the bipolar
spherical harmonics \cite{Varshalovich1988a}, which are defined as
the tensor product of two spherical harmonics
\[
{\cal Y}_{LM}^{l_{1}l_{2}}(\hat{n}_{1},\hat{n}_{2})\equiv\left[\vec{Y}_{l_{1}}(\hat{n}_{1})
\otimes\vec{Y}_{l_{1}}(\hat{n}_{2})\right]_{LM} \,,
\]
where $\vec{Y}_{l_{1}}$ is a shorthand notation for $(Y_{l_{1},-l_{1}};\dots;Y_{l_{1},+l_{1}})$.
In terms of this basis the 2pcf reads
\begin{equation}
C\left(\hat{n}_{1},\hat{n}_{2}\right)=\sum_{L,M}\sum_{l_{1},l_{2}}
{\cal A}_{l_{1}l_{2}}^{LM}{\cal Y}_{LM}^{l_{1}l_{2}}(\hat{n}_{1},\hat{n}_{2})
\,.\label{c-biposh}
\end{equation}
The coefficients ${\cal A}_{l_{1}l_{2}}^{LM}$ are called the  BipoSH
spectrum~\cite{Hajian2003a,Hajian2004,Hajian2005}. They are given by a quadratic combination of the $a_{lm}$:
\begin{equation}
{\cal A}_{l_{1}l_{2}}^{LM}
=\sum_{m_{1},m_{2}}C_{l_{1}m_{1}l_{2}m_{2}}^{LM}\left\langle a_{l_{1}m_{1}}a_{l_{2}m_{2}}\right\rangle \,,\label{biposh-coeff}
\end{equation}
where $C_{l_{1}m_{1}l_{2}m_{2}}^{LM}$ are the Clebsch-Gordan coefficients. In the case (\ref{SI}) of 
statistical isotropy they reduce to
\[
{\cal A}_{l_{1}l_{2}}^{LM}=(-1)^{l_{1}}\sqrt{2l_{1}+1}\,C_{l_{1}}\,\delta_{l_{1}l_{2}}
\delta_{ L0}\delta_{M0}\,,
\]
so that any statistically significant detection of a non-zero ${\cal A}_{l_{1}l_{2}}^{LM}$
with $L>0$ is a sign of statistical anisotropy. This property makes the BipoSH coefficients a very 
convenient null test for statistical anisotropy. 

\subsection{Anisotropies through conic modulations\label{sub:cone}}

\begin{figure}
\begin{centering}
\includegraphics[scale=0.64]{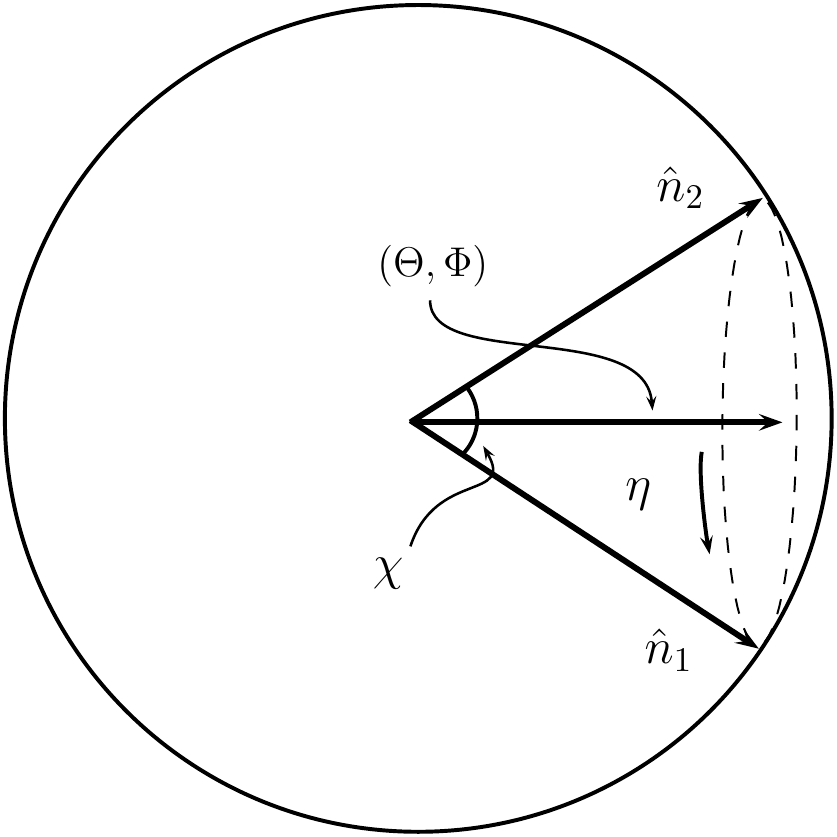}
\par\end{centering}
\caption{Two unit vectors in the CMB sphere will always define cones. The
number of angles characterizing a given cone is four, precisely the
number of degrees of freedom of the 2pcf.\label{fig:cone}}
\end{figure}

Given two unit vectors, $\hat{n}_{1}=(\chi_{1},\phi_{1})$ and $\hat{n}_{2}=(\chi_{2},\phi_{2})$, 
rooted at the origin, they define a cone on the unit sphere, obtained by rotating those vectors 
about an axis collinear to $\hat{n}_{12}\propto{\hat{n}_1+\hat{n}_2}$. Each cone can be completely 
described by three angles: the opening angle $\chi$ of the cone, and two angles $\Theta$ and $\Phi$
giving the orientation of the axis $\hat{n}_{12}$. The ordered pair of directions, 
$(\hat{n}_1,\hat{n}_2)$, is fixed by a fourth angle $\eta$ specifying their position on the circle 
bounding the intersection of the cone with the unit sphere\footnote{We adopt the convention that 
$\eta$ is the angle between the arc of the great circle from $\hat{n}_1$ to $\hat{n}_2$ and the arc 
of the great circle from $-\hat{z}$ to $\hat{z}$ through $\hat{n}_{12}$.}
 -- see Fig. \ref{fig:cone}.
Thus, instead of representing the degrees of freedom of the 2pcf
by the usual spherical angles, as in Eq. (\ref{general-2pcf}), we
can use the four angles defined by the cone:
\begin{equation}
C(\hat{n}_{1},\hat{n}_{2})=C(\chi,\eta,\Theta,\Phi)\,.
\label{cone-dof}
\end{equation}
These angles range over
\[
0\leq\Theta\leq\pi\,,\quad0\leq\Phi\leq2\pi\,,\quad
0\leq\chi\leq\pi\,,\quad0\leq\eta\leq2\pi\,.
\]
Note that by construction $\chi$ is the usual angle between $\hat{n}_1$ and $\hat{n}_2$ \cite{Varshalovich1988a}
\be
\label{scalar-chi}
\cos\chi = \cos\chi_1\cos\chi_2+\sin\chi_1\sin\chi_2\cos(\phi_1-\phi_2)\,.
\ee

The 2pcf represented by this new set of angles can be expanded in the following way
\begin{align}
\label{cone-expanded}
C(\chi,\eta,\Theta,\Phi) = &\sum_{L,M,M'}\sum_{l}\frac{(2l+1)(2L+1)}{4\pi}
{\cal C}_{l}^{LMM'}\nn \\
 & \qquad\quad\times P_{l}(\cos\chi)D_{MM'}^{L}(\eta,\Theta,\Phi),
\end{align}
where $D_{MM'}^{L}$ is the Wigner $D$-matrix, $P_{l}$ are the Legendre polynomials, 
and ${\cal C}_{l}^{LMM'}$ are the multipolar coefficients of the expansion, which we 
term the \emph{angular-conic spectrum}. We stress that the order of the angles $(\eta,\Theta,\Phi)$ 
in $D^L_{MM'}$ is important, since they are associated to the eigenvalues $M$, $L$ and $M'$, in 
this order. Note also that the exchange symmetry (\ref{exchange-symmetry})
now becomes $C(\chi,\eta,\Theta,\Phi)=C(\chi,\eta\pm\pi,\Theta,\Phi)$,
which implies that $M$ is even in the decomposition (\ref{cone-expanded}).

In order for the multipolar coefficients ${\cal C}^{LMM'}_l$ to be useful, they
need to be related to the $a_{lm}$ defined in (\ref{general-2pcf}), since the latter 
are more easily extracted from CMB maps. In principle, the relation among them can be 
found by equating Eq.~(\ref{cone-expanded}) with Eq.~(\ref{general-2pcf}) 
and by using the orthogonality of the functions $D_{MM'}^{L}(\eta,\Theta,\Phi)$ and 
$P_{l}(\cos\chi)$ to write ${\cal C}_{l}^{LMM'}$ as a function of 
$\langle a_{l_{1}m_{1}}a_{l_{2}m_{2}}\rangle$. If performed naively, however, this 
task will lead to very complicated integrals coupling the conic angles $\left(\chi,\eta,\Theta,\Phi\right)$
and the spherical angles $\left(\chi_{1},\phi_{1},\chi_{2},\phi_{2}\right)$.
The easiest way to proceed is to make use of the fact that the 2pcf is a scalar, 
and therefore the equality between Eqs.~\eqref{cone-expanded} and~\eqref{general-2pcf} 
must hold in any coordinate system. We thus specialize the decompositions (\ref{cone-expanded}) and 
(\ref{general-2pcf}) to a coordinate system in which $\Theta$, $\Phi$ and $\eta$ are zero 
but $\chi$ is not. Once the integral over $\chi$ is done, we rotate the system back to a general frame. 
This rotation can be performed using $\left(\eta,\Theta,\Phi\right)$
as the three Euler angles, which can then be moved to the right-hand
side of the equality using the orthogonality of the Wigner $D$-matrices.
The details of this computation are given in the appendix. 

The final result is
\begin{equation}
{\cal C}_{l}^{LMM'}=2\pi\sum_{\substack{l_{1},m_{1}\\
l_{2},m_{2}}}\langle a_{l_{1}m_{1}}a_{l_{2}m_{2}}\rangle\left(\begin{array}{ccc}
L & l_{1} & l_{2}\\
M' & m_{1} & m_{2}
\end{array}\right)J_{l_{1}l_{2}}^{lLM}\,,\label{conic-coeff}
\end{equation}
where $J_{l_{1}l_{2}}^{lLM}$ is a (non-square) matrix that couples
the multipoles of the conic and spherical decompositions. This matrix
comes entirely from the geometry of the problem, which means that its entries
need to be computed only once. They are defined as
\begin{equation}
J_{l_{1}l_{2}}^{lLM}=\sum_{m,m'}\left(\begin{array}{ccc}
L & l_{1} & l_{2}\\
M & m & m'
\end{array}\right)J_{l_{1}ml_{2}m'}^{l}\,,\label{JlLM}
\end{equation}
 where 
\begin{equation}
J_{l_{1}ml_{2}m'}^{l}=\eta_{l_{1},-m}\eta_{l_{2}m'}\!\!\int_{-1}^{1}P_{l}(x)P_{l_{1}}^{-m}\left(y\right)
P_{l_{2}}^{m'}\left(y\right)dx\,,\label{jlLm}
\end{equation}
and $y=\left[(x+1)/2\right]^{1/2}$. The constant coefficients $\eta_{lm}$
are defined in (\ref{eta}).

From equation (\ref{conic-coeff}) one sees that the expected values of the  ${\cal C}_{l}^{LMM'}$
relate linearly to the covariance matrix. Furthermore, in the limit of statistical isotropy, this 
relation becomes
\begin{equation}
\label{conic-SI}
{\cal C}_{l}^{LMM'}=C_{l}\delta_{L0}\delta_{M0}\delta_{M'0}\,.
\end{equation}

Similar to what happens with the BipoSH spectrum, any non-zero detection of the angular-conic spectrum 
with $L>0$ is a measure of statistical anisotropy. However, the multipole $L$ has a simpler geometrical
meaning, since it is associated with the orientation of the cone as defined in Fig. (\ref{fig:cone}). 
The multipole $L$ can be interpreted as an indication of a conic modulation of the 2pcf over the CMB sky. 
This is an important aspect of the decomposition (\ref{cone-expanded}) --  that each of its angles has a 
clear geometrical meaning. One advantage  is that it can make it simple to integrate one (or more) of 
them in a symmetric situation. The authors of Ref. \cite{Pullen2007a}, for example, consider
a power-multipole test on the one-point function: 
$C(\hat{n},\hat{n})=\left\langle \Delta T^{2}(\hat{n})\right\rangle $.
From the perspective of this work, their test is equivalent
to Eq. (\ref{cone-expanded}) with $\chi=\eta=0$.

Finally, let us mention that the angular-conic spectrum is also linearly
related to the BipoSH spectrum. Using Eqs. (\ref{biposh-coeff}) and
(\ref{conic-coeff}), together with the transformation between Clebsch-Gordan
and Wigner 3J symbols~\eqref{clebsch-3j}, we arrive at
\begin{equation}
{\cal C}_{l}^{LMM'}=2\pi\sum_{l_{1},l_{2}}\frac{(-1)^{l_{1}-l_{2}-M}}{\sqrt{2L+1}}
{\cal A}_{l_{1}l_{2}}^{L,-M'}J_{l_{1}l_{2}}^{lLM}\,.
\label{cone_biposh}
\end{equation}
We will show in Sec. \ref{sec:null-tests} how the angular-conic spectrum
can be applied in a simple null test of statistical isotropy.

\subsection{Anisotropies through planar modulations\label{sub:disc}}

Besides defining cones, two vectors $\hat{n}_{1}$ and $\hat{n}_{2}$
also define a disc. The possibility of using a disc to represent
the 2pcf was partially explored in Refs. \cite{Pereira2009,Abramo2010}.
Here we shall generalize these results to arrive at the most general
2pcf with planar symmetries.

\begin{figure}
\begin{centering}
\includegraphics[scale=0.64]{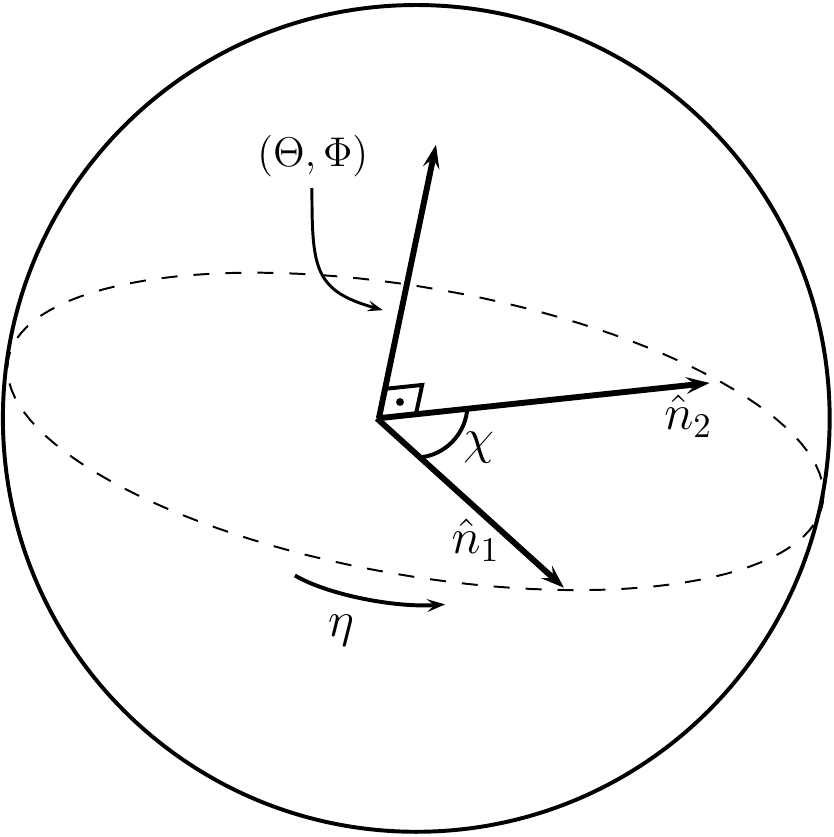}
\par\end{centering}
\caption{Two unit vectors in the CMB sphere can be used to define a disc/plane. The number
of angles characterizing this disc is three. Adding the isotropic angle $\chi$ between $\hat{n}_1$ and $\hat{n}_2$, 
we arrive at the number of angles needed to characterize the most general 2pcf.\label{fig:disc}}
\end{figure}

The geometry of the disc  is characterized by three angles: 
two angles $\Theta$ and $\Phi$ defining the overall orientation of the disc 
(i.e., its normal $\hat{N}_{12}\propto\hat{n}_{1}\times\hat{n}_{2}$)
and a third angle $\eta$ measuring the rotation of the disc around its normal\footnote{We take 
$\eta$ to be the angle from the great circle connecting the vectors 
$\hat{z}$, $-\hat{z}$ and $\hat{N}_{12}$, to the vector $\hat{n}_1$ along the disk.}
-- see Fig. (\ref{fig:disc}). 
Including finally the angle $\chi$ between $\hat{n}_{1}$ and $\hat{n}_{2}$, we have
\begin{equation}
  \label{disc-dof}
C\left(\hat{n}_{1},\hat{n}_{2}\right)=C\left(\chi,\eta,\Theta,\Phi\right)\,,
\end{equation}
where, again
\[
0\leq\Theta\leq\pi\,,\quad0\leq\Phi\leq2\pi\,,\quad0\leq\chi\leq\pi\,,\quad0\leq\eta\leq2\pi\,.
\]

As previously done for the function~\eqref{cone-dof}, the 2pcf function defined by (\ref{disc-dof}) 
can be expanded in terms of Wigner $D$-matrices and Legendre polynomials:
\begin{align}
C\left(\chi,\eta,\Theta,\Phi\right) & 
=\sum_{L,M,M'}\sum_{l}\frac{(2L+1)(2l+1)}{4\pi}{\cal D}_{l}^{LMM'}\nonumber \\
 & \qquad\quad\times P_{l}(\cos\chi) D_{MM'}^{L}(\eta,\Theta,\Phi)\,,\label{disc-expanded}
\end{align}
where ${\cal D}_{l}^{LMM'}$ are the multipolar coefficients of the expansion. The exchange symmetry 
(\ref{exchange-symmetry}) now becomes $C(\chi,\eta,\Theta,\Phi)=C(\chi,\eta,\Theta\pm\pi,\Phi)$,
which further implies that  in the decomposition (\ref{disc-expanded}) $L$ must be even. 
The decomposition~\eqref{disc-expanded} generalizes the 2pcf introduced in Refs. \cite{Pereira2009,Abramo2009}, 
where the angle $\eta$ was not included. 

Although the description so far seems to be identical with the conic
representation of the 2pcf, the relation between the disc's angles
$\left(\chi,\eta,\Theta,\Phi\right)$ and the spherical angles
$\left(\chi_{1},\phi_{1},\chi_{2},\phi_{2}\right)$ are 
different. 
Their interdependence becomes clearer when expressing the new
multipolar coefficients ${\cal D}_{l}^{LMM'}$ in terms of the covariance matrix
$\langle a_{l_{1}m_{1}}a_{l_{2}m_{2}}\rangle$.

To find this relation we choose an initial coordinate system in which
$\left(\chi,\eta,\Theta,\Phi\right)=\left(\chi,0,0,0\right)$
and $\left(\chi_{1},\phi_{1},\chi_{2},\phi_{2}\right)=\left(\pi/2,0,\pi/2,\phi_{2}\right)$.
In this coordinate system the normal to the disc points in the $z$-direction
and the remaining angle $\chi$ is equal to $\phi_{2}$ (see Eq.
(\ref{scalar-chi})). After performing the integral over $\chi$,
we rotate the system back to a general frame using $\left(\eta,\Theta,\Phi\right)$
as the three Euler angles, in that order (see the appendix for more details). 
This calculation gives
\begin{equation}
\label{planar-coeff}
{\cal D}_{l}^{LMM'}=2\pi\sum_{\substack{l_{1},m_{1}\\
l_{2},m_{2}}}\langle a_{l_{1}m_{1}}a_{l_{2}m_{2}}\rangle\left(\begin{array}{ccc}
L & l_{1} & l_{2}\\
M' & m_{1} & m_{2}
\end{array}\right)I_{l_{1}l_{2}}^{lLM}\,,
\end{equation}
where $I_{l_{1}l_{2}}^{lLM}$ is a non-square matrix coupling the two set of angles involved. 
It is defined by
\begin{equation}
I_{l_{1}l_{2}}^{lLM}\equiv\sum_{m,m'}\left(\begin{array}{ccc}
L & l_{1} & l_{2}\\
M & m & m'
\end{array}\right)I_{l_{1}ml_{2}m'}^{l}\,,\label{IlLM}
\end{equation}
where
\begin{equation}
I_{l_{1}ml_{2}m'}^{l}\equiv\lambda_{l_{1}m}\lambda_{l_{2}m'}\int_{-1}^{1}P_{l}(x)e^{im^{\prime}
\arccos 
x}dx\,,\label{ilLM}
\end{equation}
and $\lambda_{lm}$ is a set of constant coefficients defined in (\ref{lambda}).

Equation (\ref{planar-coeff}) represents the desired relation between the multipolar coefficients 
in Eq.~\eqref{disc-expanded} and the temperature multipolar coefficients. In the limit of statistical 
isotropy these coefficients become
\begin{equation}
{\cal D}_{l}^{LMM'}=C_{l}\delta_{L0}\delta_{M0}\delta_{M'0}\,,
\label{planar-SI}
\end{equation}
as expected, since the multipole $l$ measures the isotropic angular
power of the CMB.

We see here that, as it happens with the BipoSH
and angular-conic spectra, the multipolar coefficients ${\cal D}_{l}^{LMM'}$
form a legitimate null estimator of statistical isotropy, since a
measurement of any non-zero ${\cal D}_{l}^{LMM'}$ with $L>0$ is
an indication of anisotropy. Given that $l$ measures the isotropic angular
power while $L$ measures planar modulations over an isotropic sky,
the ${\cal D}_{l}^{LMM'}$ coefficients are called the \emph{angular-planar}
spectrum \cite{Pereira2009}. Note that, again, the geometrical meaning
of each angle involved in the construction of the 2pcf is clear, allowing one to easily marginalize 
over any desired degree of freedom in a symmetric situation.

The angular-planar spectrum can be directly related to the BipoSH
spectrum. Following the same computation leading to (\ref{cone_biposh})
we find
\begin{equation}
{\cal D}_{l}^{LMM'}=2\pi\sum_{l_{1},l_{2}}\frac{(-1)^{l_{1}-l_{2}-M}}{\sqrt{2L+1}}{\cal A}_{l_{1}l_{2}}^{L,-M'}I_{l_{1}l_{2}}^{lLM}\,.
\label{disc-biposh}
\end{equation}
In conclusion, the matrices $J_{l_{1}l_{2}}^{lLM}$ and $I_{l_{1}l_{2}}^{lLM}$ 
can be seen as the weights that should be added to each eigenvalue of the BipoSH 
spectrum in order to obtain the angular-conic and 
angular-planar spectra, respectively. \\

In what follows, it will be useful to introduce two new variables. Given the 
formal similarity between Eqs. (\ref{conic-coeff}) and (\ref{planar-coeff}), we will define
\begin{equation}
{\cal S}_{l}^{LMM'}\equiv2\pi\sum_{\substack{l_{1},m_{1}\\
l_{2},m_{2}}}\langle a_{l_{1}m_{1}}a_{l_{2}m_{2}}\rangle\left(\begin{array}{ccc}
L & l_{1} & l_{2}\\
M' & m_{1} & m_{2}
\end{array}\right)Q_{l_{1}l_{2}}^{lLM}\,,\label{s-coef}
\end{equation}
where ${\cal S}^{LMM'}_l$ stands for both the angular-conic and angular-planar 
power spectra, and $Q^{LMM'}_l$ represents their respective coupling 
matrices. That is
\be
{\cal S}_{l}^{LMM'}\equiv\begin{cases}
{\cal C}_{l}^{LMM'}\\
{\cal D}_{l}^{LMM'}
\end{cases}\textrm{and}\;\quad Q_{l_{1}l_{2}}^{lLM}\equiv\begin{cases}
J_{l_{1}l_{2}}^{lLM}\\
I_{l_{1}l_{2}}^{lLM}
\end{cases}.
\ee
This will allow us to put all expressions in a unified description.
\\

Before we move on, let us illustrate the use of the angular-planar and angular-conic spectra 
to distinguish between different sources of anisotropy. First, let us consider anisotropic maps
satisfying
\be
\langle a_{\ell_1m_1}a^*_{\ell_2m_2}\rangle = f(\ell_1)\delta_{\ell_1\ell_{2}\pm p}\delta_{m_1m_2}
\ee
where $f$ is some predicted function of $\ell_1$  and $p$ is any \emph{odd} integer. This is a 
simple model of parity-violating anisotropy~\cite{Abramo2010}, and can arise in many different 
theoretical contexts~\cite{Carroll:2008br,Amendola2011,Yoho01072013,Prunet:2004zy}.  For this 
covariance matrix the coefficients~\eqref{s-coef} become
\begin{align}
{\cal S}_{l}^{LM0}& =2\pi 
\sum_{\ell_1}f_{\ell_1} Q_{\ell_1\ell_1\mp p}^{lLM} \nn\\
&\times\left[\sum_{m_1}(-1)^{m_1}\left(\begin{array}{ccc}
L & \ell_1 & \ell_1\mp p\\
0 & m_1 & -m_1
\end{array}\right)\right]\,.
\end{align}
Due to momentum conservation, the quantity inside square brackets is non-zero only when $L$ is an 
odd integer. However, since the symmetry of the 2pcf restricts the planar multipole $L$ to even 
values, for this particular example we have
\be
{\cal C}^{LM0}_l\neq0\quad {\rm and}\quad {\cal D}^{LM0}_l=0\,.
\ee
Likewise, models predicting a covariance matrix of the form 
$\langle a_{\ell_1m_1}a^*_{\ell_2m_2}\rangle = f(\ell_1,m_1)\delta_{\ell_1\ell_2}\delta_{m_1m_2\pm p}$, 
such as happens with CMB in the presence of a homogeneous magnetic field~\cite{Kahniashvili:2008sh}, 
will lead to
\be
{\cal C}^{LM,\mp p}_l=0\quad {\rm and}\quad {\cal D}^{LM,\mp p}_l\neq0\,.
\ee
where we used the condition $M'\pm p = 0$ imposed by the 3J symbol. Since $M'$ has to be an even 
number for the angular-planar spectrum we conclude that ${\cal C}^{LM,\mp p}_l=0$ in this example.
Evidently, exact results as above will not hold in practice, where all sorts of statistical noise 
and foregrounds might contribute differently to each spectra. For this one has to construct 
statistical estimators from the theoretical spectra which can be directly applied to a 
given CMB map. We next discuss how such estimators can be constructed.

\section{Null tests of isotropy\label{sec:null-tests}}

An interesting feature of the angular-conic and angular-planar spectra
is that they can be used as null tests of isotropy that can potentially reveal 
the mechanisms producing the deviations from isotropy, thus giving hints on the 
mechanisms behind the observed signal. The geometrical interpretation 
of Eqs.~\eqref{cone-dof} and~\eqref{disc-dof} allows for the reduction of the 
number of angles in the 2pcf whenever the peculiarities
of the analysis permit. Most important, though, is the fact that this
feature allows us to control cosmic variance in a systematic way.
Based on that, and having the reduction of cosmic variance in mind, in this work 
we will make the simplifying assumption that the angle $\eta$ in Eqs. 
(\ref{cone-dof}) and (\ref{disc-dof}) will not lead to significant 
modulations. In other words, we work with
\begin{equation}
C(\chi,\Theta,\Phi)=\frac{1}{2\pi}\int_{0}^{2\pi}
C(\chi,\eta,\Theta,\Phi)d\eta\,,\label{simplified-2pcf}
\end{equation}
which corresponds to taking the $\eta$-monopole of Eqs. (\ref{cone-expanded})
and (\ref{disc-expanded}). Thus, from now on we shall use
\be
\label{Meq0}
M=0\,.
\ee
For convenience we can also drop the prime on $M^\prime$,
so we replace 
$M'\rightarrow M$, $S_{l}^{L0M}\rightarrow S_{l}^{LM}$ and 
$Q_{l_{1}l_{2}}^{lL0}\rightarrow Q_{l_{1}l_{2}}^{lL}$. 

Our primary motivation to assume Eq.~\eqref{Meq0} is simplicity, since it allows us to implement our method 
more easily. Nonetheless, it is
important to justify this choice geometrically. Recall that, in the case of the planar 2pcf, the angle
$\eta$ measures the rotation of the disc around its (fixed) normal. Thus, by assuming that $M=0$ we will not be 
able to detect correlations of temperature along great circles in the sky, if they exist. In the 
case of the conic 2pcf the same angle will measure correlations of temperatures over small circular rings in the sky; again,
such rings will not be detect if $M=0$. Since it is not obvious that these features lie among known anomalies, this simplification 
seems appropriate in a first analysis. However, a thorough assessment of the CMB maps with the complete 
tools presented here can potentially reveal correlations of the type we are neglecting; the results of these analyses are in progress
and shall be presented soon.
\\
 
A chi-square (null) test of conic/planar anisotropies can now be constructed. 
A simple unbiased estimator of the multipolar coefficients is
\begin{equation}
  \label{s-hat}
\widehat{{\cal S}}_{l}^{LM}\equiv2\pi\sum_{\substack{l_{1},m_{1}\\l_{2},m_{2}}}
a_{l_{1}m_{1}}a_{l_{2}m_{2}}\left(\begin{array}{ccc}
L & l_{1} & l_{2}\\
M & m_{1} & m_{2}
\end{array}\right)Q_{l_{1}l_{2}}^{lL}\,.
\end{equation}
Its covariance around some expected theoretical value, $\bar{{\cal S}}_{l}^{LM}$, 
is given by
\[
{\cal M}_{l}^{LL'MM'}=\left\langle \left(\widehat{{\cal S}}_{l}^{LM}-
\bar{{\cal S}}_{l}^{LM}\right)^{*}\left(\widehat{{\cal S}}_{l}^{L'M'}
-\bar{{\cal S}}_{l}^{L'M'}\right)\right\rangle \,.
\]
Clearly, the most interesting theoretical model to test is $\Lambda$CDM, for which 
\[
\bar{{\cal S}}_{l}^{LM}=0\,,\qquad(L>0)
\]
as follows from Eqs. (\ref{conic-SI}) and (\ref{planar-SI}). For this 
particular model, and assuming Gaussianity of the temperature fluctuations, 
Eq.~\eqref{SI} holds. Then, with the help of Wick's theorem, 
\begin{equation}
{\cal M}_{l}^{LL'MM'}=\left(\sigma_{l}^{L}\right)^{2}\delta_{LL'}\delta_{MM'}\,,\label{cov-matrix-M}
\end{equation}
where
\begin{equation}
\left(\sigma_{l}^{L}\right)^{2}\equiv\frac{8\pi^{2}}{2L+1}\sum_{l_{1},l_{2}}C_{l_{1}}C_{l_{2}}
\left(Q_{l_{1}l_{2}}^{lL}\right)^{2}\,.\label{sigmaLl}
\end{equation}
The fact that the matrix (\ref{cov-matrix-M}) is diagonal
in the $\Lambda$CDM model is a consequence
of the statistical independence of the $a_{lm}$s in this model. 
As expected, this matrix depends exclusively on the angular power spectrum
$C_{l}$, since this quantity completely
defines the statistics in $\Lambda$CDM.

Given the estimator (\ref{s-hat}) and its variance (\ref{cov-matrix-M}), we define
\begin{equation}
\label{chi-square}
\left(\chi^{2}\right)_{l}^{L}\equiv\frac{1}{2L+1}\sum_{M=-L}^{L}
\frac{\left|\widehat{{\cal S}}_{l}^{LM}\right|^{2}}
{\left(\sigma_{l}^{L}\right)^{2}}\,,
\end{equation}
which is just a chi-square test divided by the $2L+1$ conic/planar
degrees of freedom. Since by construction 
$\langle\left(\chi^2\right)^L_l\rangle=1$, we define for simplicity
\begin{equation}
\label{chi-square1}
\left(\overline{\chi}^2\right)_{l}^{L}\equiv\left(\chi^{2}\right)_{l}^{L}-1\,.
\end{equation}
Thus, any detection of $\left(\overline{\chi}^2\right)_{l}^{L}\neq0$
for $L>0$ is an indication of statistical anisotropy.

An important remark is in order. If all the data one has is a single CMB map, the 
statistics~\eqref{chi-square} should be computed entirely in terms of that 
map's data. In fact, this is the essence of a null test. Given a map, we treat 
it as if it were a $\Lambda$CDM map, and compute~\eqref{chi-square} accordingly. 
The computation of $\sigma^L_l$ (or any other piece entering 
Eq.~\eqref{chi-square}) using a set of theoretical $C_l$s, which supposedly generates
the map at hand, will cease to be a null test, and will only bias our final 
result towards {\it a priori} expectations. Thus, in the case of a single map we compute
$\sigma^L_l$ with the power spectrum estimated by $\widehat{C}_l=(2l+1)^{-1}\sum_m|a_{lm}|^2$.

\subsection{Null tests of isotropy with Planck data\label{sub:data}}

\begin{figure}
\begin{centering}
\includegraphics[scale=0.57]{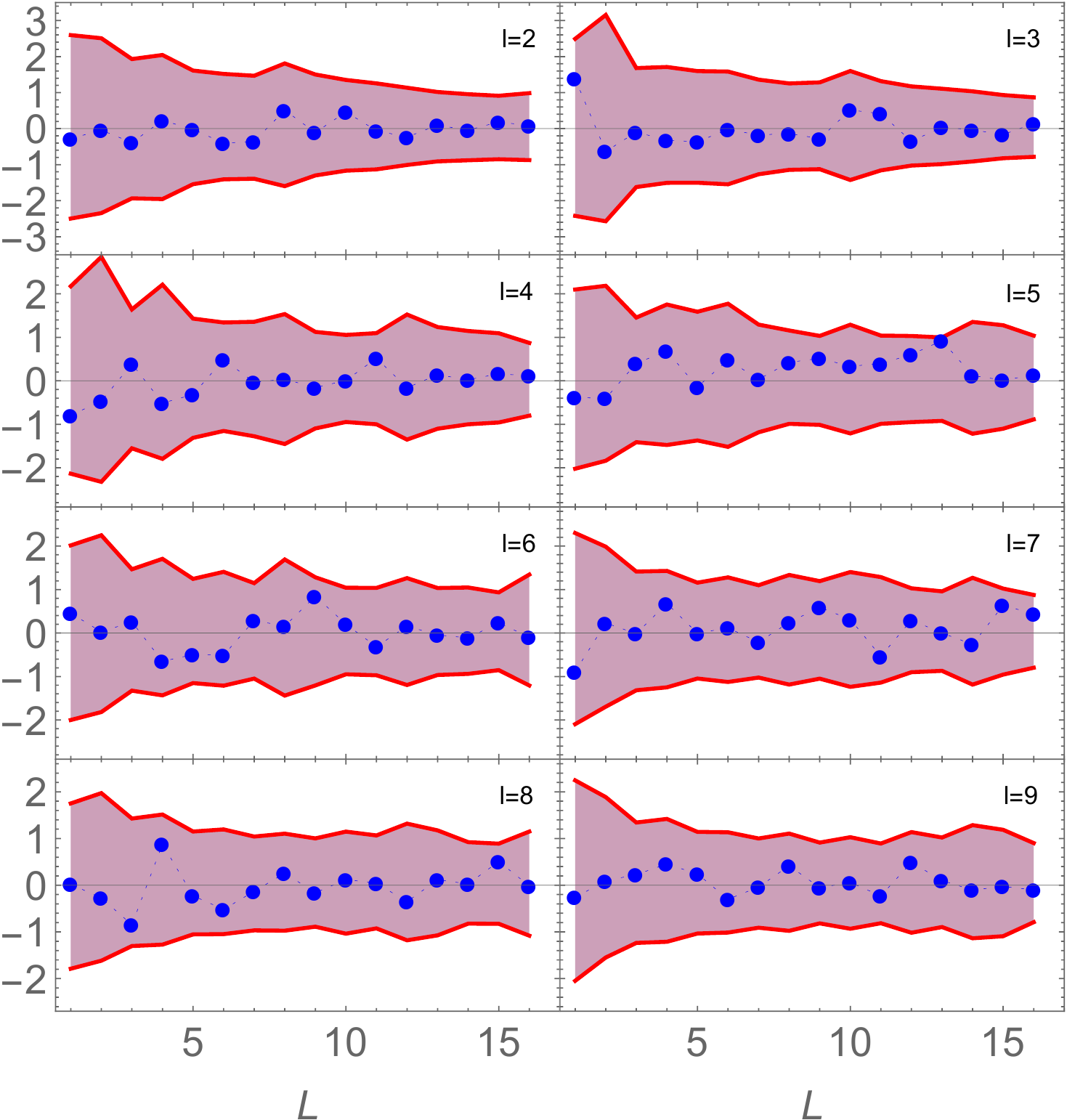}
\end{centering}
\caption{Angular-conic null test of isotropy applied to the {\it Planck} {\tt Commander} 2015 
temperature map, along with its mask. 
The plot shows the quantity $(\bar{\chi}^{2})_{l}^{L}$ versus $L$ for different angular multipoles $l$. 
The shaded contours represent $2\sigma$ cosmic variance from $10^{3}$ FFP6 simulations using 
the same mask. We also performed analyses with the other three {\it Planck} foreground-cleaned 2015 
maps: {\tt SMICA}, {\tt NILC}, and {\tt SEVEM}, obtaining qualitatively similar results.
\label{fig:cone-planck}}
\end{figure}
\begin{figure}
\begin{centering}
\includegraphics[scale=0.57]{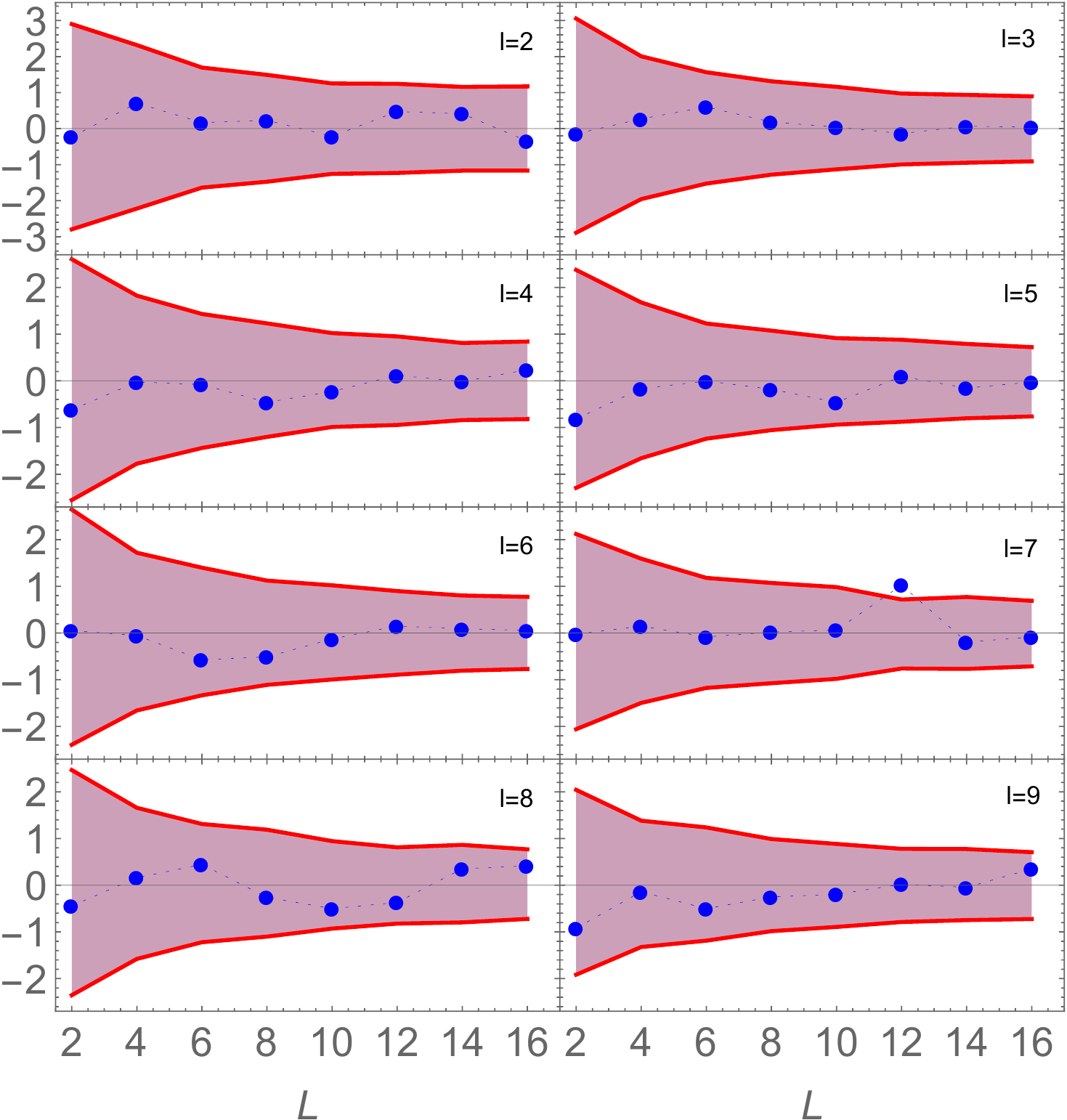}
\end{centering}
\caption{Angular-planar null test of isotropy applied to the {\it Planck} {\tt Commander} 2015 
temperature map, along with its mask. The plot shows the quantity $(\bar{\chi}^{2})_{l}^{L}$ versus 
$L$ for different angular multipoles $l$. The shaded contours represent $2\sigma$ cosmic variance 
from $10^{3}$ FFP6 simulations using the same mask. We also performed analyses with the other three 
{\it Planck} foreground-cleaned 2015 maps: {\tt SMICA}, {\tt NILC}, and {\tt SEVEM}, obtaining 
qualitatively similar results. Note that, for this test, the planar multipole $L$ is always even.
\label{fig:disc-planck}}
\end{figure}

In this section we perform a simple statistical analysis of the 2015 {\it Planck} data 
release~\cite{Ade2015a} using the tools we developed. While this analysis is not 
intended to be exhaustive, it offers an important sanity check of the available 
data. For these analyses we have used the {\it Planck} {\tt Commander} 2015 temperature map, 
along with its mask. 
The data points were compared with the 1000 Full Focal Plane (FFP6) simulations 
provided by the {\it Planck} team, to which we have also applied the {\tt Commander-Ruler} 
mask, so as to ensure that we are comparing quantities with the same foreground treatments. 
The results of our analysis are shown in Figs.~\eqref{fig:cone-planck} 
and~\eqref{fig:disc-planck}, where the data points are compared to the $2\sigma$ cosmic variance 
bars from the FFP6 simulations. We have also performed analyses with the other three {\it Planck} 
foreground-cleaned 2015 maps: {\tt SMICA}, {\tt NILC}, and {\tt SEVEM}, obtaining qualitatively 
similar results.

Our results show no drastic discrepancies between the {\it Planck} 2015 data and the
$\Lambda$CDM model at the multipoles we tested, although it is interesting to 
note that the cases $(\overline{\chi}^2)^L_{l=3}$ for the cone and 
$(\overline{\chi}^2)^L_{l=4,5}$ for the disc present a consistent deficit of 
conic/planar modulations in the scales we considered. It is important to mention 
that, while the conic/planar multipoles $L$ are statistically independent in the 
$\Lambda$CDM model 
(see Eq.~\eqref{cov-matrix-M}), the angular multipoles $l$ are not. In other words, 
while the points in each frame of Figs.~\eqref{fig:cone-planck}
and~\eqref{fig:disc-planck} are independent from each other, they are not independent
from the points in other frames of the same figure. Given that there are 16 
independent points in the multipole range we considered, on average, only $16\times4.55\%=0.73$
of them should fall outside the $2\sigma$ ($95.45\%$ Cl) variance error bars. This analysis is in 
clear agreement with our findings.

\section{Conclusions and Perspectives\label{sec:conclusions}}

The impressive agreement of the $\Lambda$CDM model with current CMB data
compel us to go beyond the simple statistical framework of a Gaussian, 
homogeneous and isotropic universe. Assuming that CMB is Gaussian, a fact which 
is supported by the latest {\it Planck} data, hints of new physics can be found in the 
realm of statistical anisotropy -- a possibility which is still open to debate.

In this work we have proposed two new representations of the 2pcf as alternatives 
to the popular bipolar power spectrum (BipoSH) analysis. Besides being model 
independent, these tools are entirely based on the geometry defined by the 
two unit vectors in the CMB sphere, namely, a cone and a disc. These tools differ 
from the BipoSH analysis in two main aspects. First, the new decompositions of 
the 2pcf are geometrically inspired, which means that null tests of isotropy 
based on their multipolar coefficients can help to elucidate the physical mechanism 
behind signals of anisotropy. We have illustrated this feature with concrete 
examples of anisotropic maps in which only one of these estimators will be non-zero.
Thus, for example, a statistically significant detection of the angular-planar spectrum
cannot result from parity-violating physics, since in this case the angular-planar spectrum is 
zero for all multipoles. Second, the clear geometrical role of the anisotropic degrees of freedom 
used as variables in the 2pcf allows us to construct simpler correlation functions whenever we have 
a more symmetric situation at hand. This feature has the important consequence of allowing us to 
reduce cosmic variance in a systematic way.

The angular-conic and angular-planar null tests of isotropy have not shown significant 
deviations of statistical isotropy in the lowest multipoles of the {\it Planck} data, although 
some angular multipoles presented interesting low conic/planar modulations. Our analysis does 
not reveal any new statistical anomaly, although known features, such as the hemispherical power 
anomaly, are expected to appear at higher angular multipoles $\l$, which were not included in our first analysis.

A more complete analysis of existing observational data, including a larger range of multipoles and a 
better assessment of systematics, is postponed to a future work. Of particular interest is the application 
of the angular-conic spectra in the investigation of the hemispherical power asymmetry found in 
Refs.~\cite{Akrami2014,Bernui08,Bernui2014}. Indeed, these references considered 
a pixel-based variance estimator which resembles in many ways the conic degrees of freedom that we 
have introduced. In this respect, it is worth mentioning that the conic 2pcf might have some 
relevance for the detection of Baryonic Acoustic Oscillations. Indeed, the ring-like pattern that 
BAO produces in the distribution of galaxies leads exactly to a three-dimensional cone centred on 
the observer.

\begin{acknowledgments}
We thank Raul Abramo and Miguel Quartin for their feedback during the development
of this work. This work was supported by the Brazilian funding agencies CNPq (Conselho 
Nacional de Desenvolvimento Científico e Tecnológico) and Capes (Coordenação de 
Aperfeiçoamento de Pessoal de Nível Superior, PVE Program, Number 88881.064966/2014-01). 
GDS is  supported by a Department of Energy grant DE--SC0009946 to the particle astrophysics theory 
group at Case Western Reserve University. Some results are based on observations obtained with {\it 
Planck} (http://www.esa.int/Planck), an ESA science mission with instruments and contributions 
directly funded by ESA Member States, NASA, and Canada, which we acknowledge. 
\end{acknowledgments}

\appendix

\section{}

We present here the derivation of our main results, Eqs.~\eqref{conic-coeff} 
and~\eqref{planar-coeff}. We also collect useful identities and mathematical 
formulas used in the text.

\subsection{Spherical Harmonics and Wigner 3J symbols}

Our definition for the spherical harmonics is
\[
Y_{lm}(\chi,\phi)=(-1)^{m}\eta_{lm}P_{l}^{m}(\cos\chi)e^{im\phi}\,,
\]
where
\begin{equation}
\eta_{lm}=\sqrt{\frac{2l+1}{4\pi}\frac{(l-m)!}{\left(l+m\right)!}}=\eta_{l,-m}\frac{(l-m)!}{(l+m)!}\,.\label{eta}
\end{equation}
At the point $(\chi,\phi)=(\pi/2,\phi)$, it simplifies to 
\be
Y_{lm}(\pi/2,\phi)=\lambda_{lm}e^{im\phi}\,,
\ee
where
\begin{align}
\lambda_{lm}& = (-1)^{\frac{(l+m)}{2}}\times\nn \\
 & \begin{cases}
\!\left[\!\frac{2l+1}{4\pi}\frac{(l+m-1)!!}{(l+m)!!}\frac{(l-m-1)!!}{(l-m)!!}\!\right]^{1/2}\!\! & l+m=\textrm{even}\\
0 & \textrm{otherwise}\,.
\end{cases}\label{lambda}
\end{align}

The relation between Wigner 3Js and Clebsch-Gordan coefficients are
\be
\label{clebsch-3j}
\left(\begin{array}{ccc}
l_{1} & l_{2} & l_{3}\\
m_{1} & m_{2} & -m_{3}
\end{array}\right)=\frac{(-1)^{l_{1}-l_{2}-m_{3}}}{\sqrt{2l_{3}+1}}C_{l_{1}m_{1}l_{2}m_{2}}^{l_{3}m_{3}}\,.
\ee
Other useful identities include
\begin{align}
\label{useful-eq1}
\sum_{m}(-1)^{l-m}\left(\begin{array}{ccc}
L & l & l\\
0 & m & -m
\end{array}\right) & =\sqrt{2l+1}\delta_{L0}\,,\\
\label{useful-eq2}
\left(\begin{array}{ccc}
0 & l & l\\
0 & m & -m
\end{array}\right) & =\frac{(-1)^{l-m}}{\sqrt{2l+1}}\,.
\end{align}

We also remind two useful orthogonality relations of the Wigner rotation matrices. These are
\[
\int D^{l_{1}}_{m_{1}m'_{1}}(w)D^{l_{2}*}_{m_{2}m'_{2}}(w)dw=
8\pi^2\delta_{\ell_1\ell_2}\delta_{m_1m_2}\delta_{m_1'm_2'}\,,
\]
and
\begin{align*}
\int D^{l_{1}}_{m_{1}m'_{1}}(w)&D^{l_{2}}_{m_{2}m'_{2}}(w)D^{l_{3}}_{m_{3}m'_{3}}(w)dw=
\\
&8\pi^2\left(\begin{array}{ccc}
l_{1} & l_{2}  & l_{3} \\
m'_{1}  & m'_{2} & m'_{3}
\end{array}\right)\left(\begin{array}{ccc}
l_{1} & l_{2} & l_{3}\\
m_{1} & m_{2} & m_{3}
\end{array}\right)\,,
\end{align*}
with $w=(\alpha,\beta,\gamma)$ being the three Euler angles.

\subsection{Derivation of Eq. (\ref{conic-coeff}) for conic-angular modulations}

After expanding the 2pcf in terms of Legendre polynomials and Wigner $D$-matrices, we equate expressions~(\ref{cone-expanded}) 
and~(\ref{general-2pcf}). Since the 2pcf is a scalar, this equality should hold in a coordinate system in which the symmetry 
axis of the cone is aligned with the $z$-axis. That will mean:
\[
\Theta=\Phi=\eta=0\,,\quad\chi_{1}=\chi_{2}=\chi/2\,,\quad\phi_{1}=\phi_{2}-\pi=0\,.
\]
Using the identity $D_{MM'}^{L}(0,0,0)=\delta_{MM'}$ and the orthogonality
of the Legendre polynomials, we then arrive at
\begin{equation}
\sum_{L,M}\frac{(2L+1)}{2\pi}{\cal C}_{l}^{LMM}=\sum_{\substack{l_{1},m_{1}\\
l_{2},m_{2}}
}\left\langle a_{l_{1}m_{1}}a_{l_{2}m_{2}}\right\rangle 
J_{l_{1}m_{1}l_{2}m_{2}}^{l}\,,\label{cone-part1}
\end{equation}
where $J_{l_{1}m_{1}l_{2}m_{2}}^{l}$ was defined in Eq. (\ref{jlLm}).
If we now rotate the axes back to a general coordinate system using $\omega=\left\{ \eta,\Theta,\Phi\right\}$ as 
the three Euler angles, the coefficients ${\cal C}_{l}^{LMM}$ and $a_{lm}$ will transform as \cite{Varshalovich1988a}
\[
{\cal C}_{l}^{LMM}\!=\!\sum_{M'}\tilde{{\cal C}}_{l}^{LMM'}D_{MM'}^{L}(\omega),\quad 
a_{lm}\!=\!\sum_{m'}\tilde{a}_{lm'}D_{mm'}^{l}(\omega)\,.
\]
Then, we multiply both sides of (\ref{cone-part1}) by $D_{M'M''}^{L'}(\omega)$
and use the orthogonality of the Wigner rotation matrices to isolate
$\tilde{{\cal C}}_{l}^{LMM'}$. Finally, we use $(-1)^{M+M'}\tilde{{\cal C}}_{l}^{L,-M,-M'}=\tilde{{\cal C}}_{l}^{LMM'*}$, 
which follows from the reality of the 2pcf, and substitute ${\cal C}_{l}^{LMM'*}\rightarrow{\cal C}_{l}^{LMM'}$,
since we could equally well have started with the complex conjugate
in the expansion (\ref{cone-expanded}). Dropping primes and tildes, we arrive at
(\ref{conic-coeff}).

\subsubsection{Isotropic limit}

In order to derive the isotropic limit~\eqref{conic-SI}, we note that, in this limit, 
$\langle a_{\ell_1m_1}a_{\ell_2m_2}\rangle = 
(-1)^{m_2}C_{\ell_1}\delta_{\ell_1\ell_2}\delta_{m_1,-m_2}$. Since the Wigner 3J symbol 
appearing in Eq.~\eqref{conic-coeff} is zero unless $M'+m_1+m_2=0$, this implies that $M'=0$. Then, 
using Eq.~\eqref{useful-eq1}, we arrive at
\be\label{cone-isolim}
{\cal C}^{LMM'}_l = 
2\pi\sum_{l_1}C_{l_1}(-1)^{l_1}\sqrt{2l_1+1}J^{lLM}_{l_1l_1}\times\delta_{L0}\delta_{M'0}\,.
\ee
In the above expression, the non-vanishing terms of the coupling matrix are of the form 
$J^{l0M}_{l_1l_2}$. Combining Eq.~\eqref{useful-eq2} with the addition theorem for the associated 
Legendre polynomials one can show that
\be
J^{l0M}_{l_1l_2} = \frac{(-1)^{l_1}}{\sqrt{2l_1+1}}\frac{1}{2\pi}\delta_{ll_1}\delta_{M0}\,.
\ee
Combining this result with~\eqref{cone-isolim}, we finally get~\eqref{conic-SI}.

\subsection{Derivation of Eq. (\ref{planar-coeff}) for planar-angular modulations}

Equating the expanded 2pcf (\ref{disc-expanded}) with (\ref{general-2pcf}), we choose a particular coordinate system in 
which the normal to the disc points in the $z$-direction, which means:
\[
\Theta=\Phi=\eta=0,\quad\chi_{1}=\chi_{2}=\pi/2,\quad\phi_{1}=0,\quad\chi=\phi_{2}\,.
\]
Using $D_{M'M}^{L}(0,0,0)=\delta_{M'M}$, $Y_{lm}(\pi/2,\phi)=\lambda_{lm}e^{im\phi}$,
and integrating over $P_{l}(\cos\chi)$, we find
\[
\sum_{L,M}\frac{(2L+1)}{2\pi}{\cal D}_{l}^{LMM}=\sum_{\substack{l_{1},m_{1}\\
l_{2},m_{2}
}
}\left\langle a_{l_{1}m_{1}}a_{l_{2}m_{2}}\right\rangle I_{l_{1}m_{1}l_{2}m_{2}}^{l}
\]
where $I_{l_{1}m_{1}l_{2}m_{2}}^{l}$ was introduced in (\ref{ilLM}).
We now rotate back to a general coordinate system using $\omega=\left\{ \eta,\Theta,\Phi\right\} $
as the Euler angles and the fact that, under rotations
\[
{\cal D}_{l}^{LMM}\!=\!\sum_{M'}\tilde{{\cal D}}_{l}^{LMM'}D_{MM'}^{L}(\omega),\quad a_{lm}\!=\!\sum_{m'}\tilde{a}_{lm'}D_{mm'}^{l}(\omega)\,.
\]
From this point on, the deduction is similar to the case of the cone.
After some redefinitions and relabeling of the indices, we finally
arrive at (\ref{planar-coeff}).

\subsubsection{Isotropic limit}

The derivation of \eqref{planar-SI} follows a similar deduction to the one of the 
angular-conic power spectrum. In this limit, the expression
$\langle a_{\ell_1m_1}a_{\ell_2m_2}\rangle = 
(-1)^{m_2}C_{\ell_1}\delta_{\ell_1\ell_2}\delta_{m_1,-m_2}$ implies that $M'=0$. Then, 
using Eq.~\eqref{useful-eq1}, we arrive at
\be\label{disc-isolim}
{\cal D}^{LMM'}_l = 
2\pi\sum_{l_1}C_{l_1}(-1)^{l_1}\sqrt{2l_1+1}I^{lLM}_{l_1l_1}\times\delta_{L0}\delta_{M'0}\,.
\ee
Using again Eq.~\eqref{useful-eq2}, it is not difficult to show that
\be
I^{l0M}_{l_1l_2} = \frac{(-1)^{l_1}}{\sqrt{2l_1+1}}\frac{1}{2\pi}\delta_{ll_1}\delta_{M0}\,.
\ee
Combining this result with~\eqref{disc-isolim}, we finally get~\eqref{planar-SI}.

\bibliographystyle{h-physrev4}
\phantomsection\addcontentsline{toc}{section}{\refname}\bibliography{2pcf}

\begin{thebibliography}{10}

\bibitem{Ade2014d}
Planck, P.~Ade {\em et~al.},
\newblock Astron.Astrophys. {\bf 571}, A16 (2014), [1303.5076].

\bibitem{Ade2015a}
Planck, P.~Ade {\em et~al.},
\newblock arXiv:1502.01589.

\bibitem{Laureijs2011}
EUCLID Collaboration, R.~Laureijs {\em et~al.},
\newblock arXiv:1110.3193.

\bibitem{Abate2012}
LSST Dark Energy Science Collaboration, A.~Abate {\em et~al.},
\newblock arXiv:1211.0310.

\bibitem{Benitez2014}
J-PAS Collaboration, N.~Benitez {\em et~al.},
\newblock arXiv:1403.5237.

\bibitem{Cornish2004}
N.~J. Cornish, D.~N. Spergel, G.~D. Starkman and E.~Komatsu,
\newblock Phys.Rev.Lett. {\bf 92}, 201302 (2004), [astro-ph/0310233].

\bibitem{Kunz2006}
M.~Kunz {\em et~al.},
\newblock Phys.Rev. {\bf D73}, 023511 (2006), [astro-ph/0510164].

\bibitem{Ade2015}
Planck Collaboration, P.~Ade {\em et~al.},
\newblock arXiv:1502.01593.

\bibitem{Fabre:2013wia}
O.~Fabre, S.~Prunet and J.-P. Uzan,
\newblock arXiv:1311.3509.

\bibitem{Schmidt:2012ky}
F.~Schmidt and L.~Hui,
\newblock Phys.Rev.Lett. {\bf 110}, 011301 (2013), [1210.2965].

\bibitem{Komatsu2003}
WMAP Collaboration, E.~Komatsu {\em et~al.},
\newblock Astrophys.J.Suppl. {\bf 148}, 119 (2003), [astro-ph/0302223].

\bibitem{Ade2014c}
Planck, P.~Ade {\em et~al.},
\newblock Astron.Astrophys. {\bf 571}, A24 (2014), [1303.5084].

\bibitem{Ade2015b}
Planck Collaboration, P.~Ade {\em et~al.},
\newblock arXiv:1502.01592.

\bibitem{Bouchet2011}
COrE Collaboration, F.~Bouchet {\em et~al.},
\newblock arXiv:1102.2181.

\bibitem{Copi2010}
C.~J. Copi, D.~Huterer, D.~J. Schwarz and G.~D. Starkman,
\newblock Adv.Astron. {\bf 2010}, 847541 (2010), [1004.5602].

\bibitem{Pitrou2008}
C.~Pitrou, T.~S. Pereira and J.-P. Uzan,
\newblock JCAP {\bf 0804}, 004 (2008), [0801.3596].

\bibitem{Hanson2010a}
D.~Hanson, A.~Lewis and A.~Challinor,
\newblock Phys.Rev. {\bf D81}, 103003 (2010), [1003.0198].

\bibitem{Bennett2011}
C.~Bennett {\em et~al.},
\newblock Astrophys.J.Suppl. {\bf 192}, 17 (2011), [1001.4758].

\bibitem{Yoho01072013}
A.~Yoho, C.~J. Copi, G.~D. Starkman and T.~S. Pereira,
\newblock Mon.Not.Roy.Astron.Soc. {\bf 432}, 2208 (2013).

\bibitem{Ade2014b}
Planck, P.~Ade {\em et~al.},
\newblock Astron.Astrophys. {\bf 571}, A23 (2014), [1303.5083].

\bibitem{Amendola2011}
L.~Amendola {\em et~al.},
\newblock JCAP {\bf 1107}, 027 (2011), [1008.1183].

\bibitem{Kahniashvili2008a}
T.~Kahniashvili, G.~Lavrelashvili and B.~Ratra,
\newblock Phys.Rev. {\bf D78}, 063012 (2008), [0807.4239].

\bibitem{BH08}
A.~Bernui and W.~Hipolito-Ricaldi,
\newblock Mon.Not.Roy.Astron.Soc. {\bf 389}, 1453 (2008), [0807.1076].

\bibitem{Durrer1998}
R.~Durrer, T.~Kahniashvili and A.~Yates,
\newblock Phys.Rev. {\bf D58}, 123004 (1998), [astro-ph/9807089].

\bibitem{Hinshaw:1996ut}
G.~Hinshaw {\em et~al.},
\newblock Astrophys.J. {\bf 464}, L25 (1996), [astro-ph/9601061].

\bibitem{Spergel2003}
WMAP, D.~Spergel {\em et~al.},
\newblock Astrophys.J.Suppl. {\bf 148}, 175 (2003), [astro-ph/0302209].

\bibitem{Copi:2006tu}
C.~J. Copi, D.~Huterer, D.~J. Schwarz and G.~D. Starkman,
\newblock Phys.Rev. {\bf D75}, 023507 (2007), [astro-ph/0605135].

\bibitem{Copi:2008hw}
C.~J. Copi, D.~Huterer, D.~J. Schwarz and G.~D. Starkman,
\newblock Mon.Not.Roy.Astron.Soc. {\bf 399}, 295 (2009), [0808.3767].

\bibitem{Copi:2010na}
C.~J. Copi, D.~Huterer, D.~J. Schwarz and G.~D. Starkman,
\newblock Adv.Astron. {\bf 2010}, 847541 (2010), [1004.5602].

\bibitem{Copi:2013cya}
C.~J. Copi, D.~Huterer, D.~J. Schwarz and G.~D. Starkman,
\newblock arXiv:1310.3831.

\bibitem{Hajian:2007pi}
A.~Hajian,
\newblock astro-ph/0702723.

\bibitem{Bunn:2008zd}
E.~F. Bunn and A.~Bourdon,
\newblock Phys.Rev. {\bf D78}, 123509 (2008), [0808.0341].

\bibitem{Copi:2013zja}
C.~Copi, D.~Huterer, D.~Schwarz and G.~Starkman,
\newblock Mon.Not.Roy.Astron.Soc. {\bf 434}, 3590 (2013), [1303.4786].

\bibitem{Yoho:2013tta}
A.~Yoho, C.~Copi, G.~Starkman and A.~Kosowsky,
\newblock Mon.Not.Roy.Astron.Soc. {\bf 442}, 2392 (2014), [1310.7603].

\bibitem{Yoho:2015bla}
A.~Yoho, S.~Aiola, C.~J. Copi, A.~Kosowsky and G.~D. Starkman,
\newblock arXiv:1503.05928.

\bibitem{Varshalovich1988a}
D.~A. Varshalovich, A.~N. Moskalev and V.~K. Khersonskii,
\newblock {\em Quantum theory of angular momentum} (World Scientific, 1998).

\bibitem{Hajian2003a}
A.~Hajian and T.~Souradeep,
\newblock Astrophys.J. {\bf 597}, L5 (2003), [astro-ph/0308001].

\bibitem{Hajian2004}
A.~Hajian, T.~Souradeep and N.~J. Cornish,
\newblock Astrophys.J. {\bf 618}, L63 (2004), [astro-ph/0406354].

\bibitem{Hajian2005}
A.~Hajian and T.~Souradeep,
\newblock astro-ph/0501001.

\bibitem{Joshi2010}
N.~Joshi, S.~Jhingan, T.~Souradeep and A.~Hajian,
\newblock Phys.Rev. {\bf D81}, 083012 (2010), [0912.3217].

\bibitem{Book2012}
L.~G. Book, M.~Kamionkowski and T.~Souradeep,
\newblock Phys.Rev. {\bf D85}, 023010 (2012), [1109.2910].

\bibitem{Kamionkowski2011}
M.~Kamionkowski and T.~Souradeep,
\newblock Phys.Rev. {\bf D83}, 027301 (2011), [1010.4304].

\bibitem{Hajian2006}
A.~Hajian and T.~Souradeep,
\newblock Phys.Rev. {\bf D74}, 123521 (2006), [astro-ph/0607153].

\bibitem{Kumar2014}
S.~Kumar {\em et~al.},
\newblock arXiv:1409.4886.

\bibitem{Pullen2007a}
A.~R. Pullen and M.~Kamionkowski,
\newblock Phys.Rev. {\bf D76}, 103529 (2007), [0709.1144].

\bibitem{Pereira2009}
T.~S. Pereira and L.~R. Abramo,
\newblock Phys.Rev. {\bf D80}, 063525 (2009), [0907.2340].

\bibitem{Abramo2010}
L.~R. Abramo and T.~S. Pereira,
\newblock Adv.Astron. {\bf 2010}, 378203 (2010), [1002.3173].

\bibitem{Abramo2009}
L.~R. Abramo, A.~Bernui and T.~S. Pereira,
\newblock JCAP {\bf 0912}, 013 (2009), [0909.5395].

\bibitem{Carroll:2008br}
S.~M. Carroll, C.-Y. Tseng and M.~B. Wise,
\newblock Phys. Rev. {\bf D81}, 083501 (2010), [0811.1086].

\bibitem{Prunet:2004zy}
S.~Prunet, J.-P. Uzan, F.~Bernardeau and T.~Brunier,
\newblock Phys. Rev. {\bf D71}, 083508 (2005), [astro-ph/0406364].

\bibitem{Kahniashvili:2008sh}
T.~Kahniashvili, G.~Lavrelashvili and B.~Ratra,
\newblock Phys. Rev. {\bf D78}, 063012 (2008), [0807.4239].

\bibitem{Akrami2014}
Y.~Akrami {\em et~al.},
\newblock Astrophys.J. {\bf 784}, L42 (2014), [1402.0870].

\bibitem{Bernui08}
A.~Bernui,
\newblock Phys.Rev. {\bf D78}, 063531 (2008), [0809.0934].

\bibitem{Bernui2014}
A.~Bernui, A.~Oliveira and T.~S. Pereira,
\newblock JCAP {\bf 1410}, 041 (2014), [1404.2936].

\end{thebibliography}

\end{document}